\documentclass{ws-procs9x6}

\begin{document}

\title{BPS MAXWELL-CHERN-SIMONS-LIKE VORTICES IN A LORENTZ-VIOLATING
FRAMEWORK}
\author{R.\ CASANA,$^*$ M.M.\ FERREIRA JR., E.\ DA HORA, and A.B.F.\ NEVES}

\address{Departamento de F\'{\i}sica, Universidade Federal do Maranh\~{a}o\\
S\~{a}o Lu\'{\i}s, Maranh\~{a}o 65080-805,  Brazil\\
$^*$E-mail: rodolfo.casana@gmail.com}
\begin{abstract}
We have analyzed Maxwell-Chern-Simons-Higgs BPS vortices in a
Lorentz-violating CPT-odd context. The Lorentz violation induces profiles
with a conical behavior at the origin. For some combination of the
coefficients for Lorentz violation there always exists a sufficiently large
winding number for which the magnetic field flips its sign.
\end{abstract}

\bodymatter

\phantom{x} \vskip 10pt

\noindent Lorentz-violating (LV) theories have caught much attention since
the proposal of the Standard-Model Extension (SME) \cite{Colladay}, whose
properties have been intensively scrutinized in many areas of contemporary
physics. The study of topological defects in LV scenarios was conducted
initially for kinks \cite{Defects}, monopoles \cite{Monopole1}. Also, the
formation of topological defects (monopoles) was analyzed in a broader
framework of field theories endowed with tensor fields that spontaneously
break Lorentz symmetry \cite{Seifert}. Recently, the existence of BPS vortex
of type Abrikosov-Nielsen-Olesen has been shown in the presence of CPT-even
LV terms of the SME \cite{Carlisson1}.

The aim of this contribution is to study Chern-Simons-like BPS vortices in
an LV framework attained via the dimensional reduction of the
Maxwell-Carroll-Field-Jackiw-Higgs model \cite{EPJC1}. Such dimensional
reduction provides a Maxwell-Chern-Simons-Higgs (MCSH) model modified by LV
terms,
\begin{eqnarray}
\mathcal{L} & =&-\frac{1}{4}F_{\mu\nu}F^{\mu\nu}+\frac{1}{4}s\,\epsilon
^{\nu\rho\sigma}A_{\nu}F_{\rho\sigma}+|\partial_{\mu}\phi-ieA_{\mu}\phi
|^{2}+\frac{1}{2}\partial_{\mu}\psi\partial^{\mu}\psi  \nonumber \\
&& -e^{2}\psi^{2}|\phi|^{2}-U(|\phi|,\psi)-\frac{1}{2}\epsilon^{\mu\rho%
\sigma }(k_{AF})_{\mu}\left(
A_{\rho}\partial_{\sigma}\psi-\psi\partial_{\rho }A_{\sigma}\right) ,
\label{L2}
\end{eqnarray}
where the LV parameter $s$ plays the role of a Chern-Simons coupling, and $%
\left( k_{AF}\right) _{\mu}$ is the (1+2)-dimensional Carroll-Field-Jackiw
(CFJ) background. The potential $U\left( \left\vert \phi\right\vert
,\psi\right) =(ev^{2}-e\left\vert \phi\right\vert ^{2}-s\psi)^{2}/2$
provides the BPS vortices. The corresponding Gauss law is
\begin{equation}
\partial_{j}\partial_{j}A_{0}-sB-\epsilon_{ij}\left( k_{AF}\right)
_{i}\partial_{j}\psi=2e^{2}A_{0}\left\vert \phi\right\vert ^{2}.
\label{Gauss_20}
\end{equation}

To explicitly attain the vortex solutions,
we use the ansatz $\phi =vg\left( r\right) e^{in\theta },~A_{\theta
}=-(a(r)-n)/er,~A_{0}=\omega (r), ~\psi =\mp\omega (r)$. The functions $a(r)$, $g(r)$ and $\omega
(r)$ satisfy the following boundary conditions:
\begin{eqnarray}
&\displaystyle{g(0)=0,\,a(0)=n,\,\omega ^{\prime }(0)=\pm \omega
_{0}(k_{AF})_{\theta }/2,\,}&  \label{bc1} \\
&\displaystyle{g(\infty )=1,\,a(\infty )=0,\,\omega (\infty )=0,}&
\label{bc2}
\end{eqnarray}%
where $n$ is the winding number of the vortex configuration,  $\left( k_{AF}\right) _{\theta }$ is 
the angular component of $\left(k_{AF}\right) _{i}$ whereas the constant $\omega _{0}=\omega \left( 0\right)
$ depends on the boundary conditions and it is numerically determined. In
this ansatz the magnetic field is given by $B=-a^{\prime }/er$. \ Then, the
stationary energy density associated with the model (\ref{L2}) is written as
\begin{equation}
\mathcal{E}=\frac{1}{2}\left[ B\mp \left( ev^{2}\left( 1-g^{2}\right) \pm
s\omega \right) \right] ^{2}+v^{2}\left( g^{\prime }\mp \frac{ag}{r}\right)
^{2}\pm ev^{2}B+\partial _{a}\mathcal{J}_{a},  \label{E_den}
\end{equation}%
where the condition $\left( k_{AF}\right) _{0}=0$ was adopted to insure
positiveness of the energy density. The energy is minimized by requiring
\begin{equation}
g^{\prime }=\pm \frac{ag}{r},~{B=\pm ev^{2}\left( 1-g^{2}\right) +s\omega ,}
\label{BPS_cart2}
\end{equation}%
which are the BPS equations, where the upper (lower) sign corresponds to $n>0
$ $(n<0)$. These and the Gauss law now written as
\begin{equation}
\omega ^{\prime \prime }+\frac{\omega ^{\prime }}{r}-sB\mp (k_{AF})_{\theta
}\left( \omega ^{\prime }+\frac{\omega }{2r}\right) -2e^{2}v^{2}g^{2}\omega
=0  \label{Gauss_2}
\end{equation}%
describe topological vortices in this LV MCSH model. For $\left(
k_{AF}\right) _{\theta }=0$ one recovers the MCSH system. Under BPS
equations and boundary conditions (\ref{bc1})-(\ref{bc2}), the Eq. (\ref{E_den}) provides the BPS energy
$E_{BPS}=\pm 2\pi v^{2}n$, which is proportional to the quantized magnetic
flux, $\Phi _{B}=2\pi n/e$.

\begin{figure}
\hspace{-0cm}\psfig{file=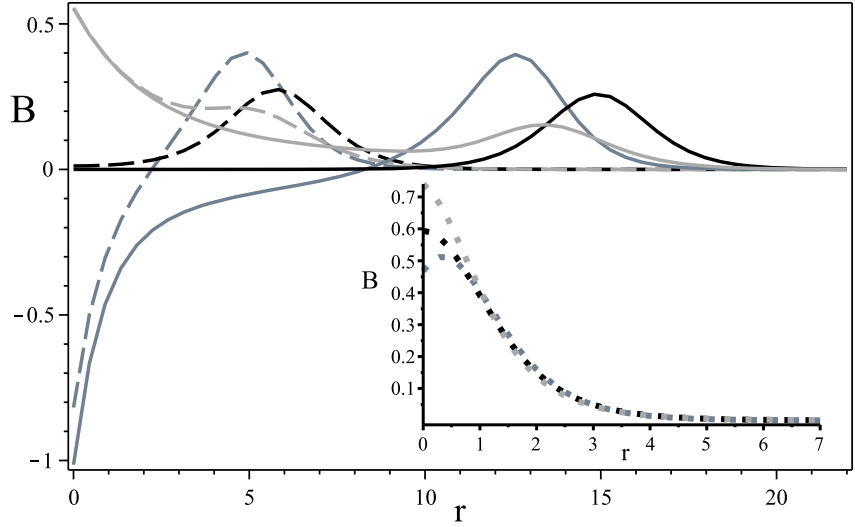,width=4.5in}
\caption{Magnetic field ${B}(r)$. Here $s=1$, $e=v=1$. Dark
gray lines, $\protect(k_{AF})_{\theta }=-1$; black lines, $\protect(k_{AF})_{\theta }=0$, MCSH;
slate gray lines, $\protect(k_{AF})_{\theta } =+1$; dotted lines, $n=1$; dashed lines, $%
n=6$; and solid lines, $n=15$.}
\label{BPS}
\end{figure}

Figure \ref{BPS} depicts  the profiles of the magnetic field obtained by numerical integration of
Eqs. (\ref{BPS_cart2})-(\ref{Gauss_2}) for $s=1$, and some values of $(k_{AF})_{\theta } $ and $n$.
For large values of radial coordinate, the profiles behave in a similar way
to the MCSH ones but with mass scales $(\beta )$ satisfying $\beta
_{{\!(k_{{AF}})_{\theta }>0}}\!\!<\beta _{{\!(k_{{AF}})_{\theta } =0}}\!\!<\beta
_{{\!(k_{{AF}})_{\theta } <0}}$. The role played by the LV parameter becomes more
pronounced at the origin, where it generates a conical structure absent in
the MCSH profiles. The numerical analysis shows that for $n>0$ (analogous
results are obtained for $n<0$), a fixed $s$, and $(k_{AF})_{\theta } >0,$
there are always two well defined regions with positive magnetic flux,
occurring no magnetic field reversion. On the other hand, for fixed $s$,
and $(k_{AF})_{\theta } <0,$ there always exists a sufficiently large winding number $%
n_{0}$\ such that for all $n>n_{0}$ the magnetic field reverses its sign.
Consequently, there are always two well defined regions with opposite
magnetic flux. The flipping of the magnetic flux represents another
remarkable feature induced by Lorentz violation.

\section*{Acknowledgments}

The authors thank FAPEMA, CNPq and CAPES for financial support.

\end{document}